# A Regional Analysis of Electric LDV Portfolio Choices by Vehicle Manufacturers

Aditya Ramji [1], Hanif Tayarani [1]

[1] *Institute of Transportation Studies, University of California at Davis*

*1605 Tilia Street, Davis, CA, 95616*

*Email: htayarani@ucdavis.edu*

**Executive Summary**

Global light duty electric vehicle (EV) sales exceeded 10.5 million units in 2022, with a year-on-year growth of 55%, but these trends differ regionally. Despite the robust growth, upfront purchase price remains a challenge for consumers in different regions, and thus, OEMs make technology choices to respond to market needs. This paper examines the electrification portfolio choices of three major automotive manufacturers (OEMs) in different regions of the world, including Europe, Americas, Asia Pacific, and Africa/Middle-East. The analysis focuses on trends in dominant segments for Battery Electric Vehicles (BEV) and Plug-in Hybrid Electric Vehicles (PHEV), as well as battery chemistry choices. Regional differences show a trend towards SUVs for both BEVs and PHEVs. Tesla's dominance in the BEV market influences battery chemistry choices. Average battery sizes for BEVs remain similar in Europe and Americas, but lower in Asia Pacific and Africa/Middle East.

## 1 Introduction

Global light duty EV sales crossed 10.5 million in 2022, growing 55% year-on-year [1]. The EVs include both Battery Electric Vehicles (BEV) and Plug-in Hybrid Electric Vehicles (PHEV). While global growth has been robust, the extent and rate of electrification differ by region, and within that by country. The Asia Pacific contributed about 63% of global EV light duty sales, essentially coming from China, followed by about 26% from Europe, 11% from the Americas region, and less than 1% from Africa, the Middle East and other low-income countries.

As these markets transition to EVs, the upfront purchase price remains a challenge for different sections of consumers across all regions [2]. It is expected that as certain markets scale, the benefits of technology cost reductions can be passed on to global markets. A key component of the cost of an EV is the battery, which is currently dominated by lithium-ion-based chemistries. Various automotive manufacturers (OEMs) have evolved over time, choosing different battery chemistries based on cost considerations, range requirements, supply constraints and geopolitics.

The average battery size per vehicle (kWh, weighted by sales) has been increasing over the years, going from about 22.2 kWh and 16.5 kWh per vehicle in 2011 for BEV and PHEV, respectively, to 61.8 kWh and 20.1 kWh for BEV and PHEV respectively in 2023[3]. Between 2010 – 2022, NMC cells (Nickel Manganese Cobalt) have been the majority. In the period up to late 2020, NCA cells (Nickel Cobalt Aluminium) were the second most preferred. In the last couple of years, LFP cells (Lithium Ferro-Phosphate) has made a resurgence with changing industry dynamics [3].

## 1.1 Research Objectives and Methodology

This paper aims to understand the choices by three key automotive manufacturers (OEMs) in terms of their LDV electrification portfolio by region, given the differing dynamics of electric vehicle adoption across countries. For the purpose of this paper, we divide the world into four regions: (i) Europe; (ii) Americas; (iii) Asia Pacific; and, (iv) Africa and Middle-East.

The analysis focuses on two research objectives. First, to understand the trends in dominant electric LDV segments, and second, to understand the trends in dominant battery chemistry choices. Both of these are analyzed from the lens of Battery Electric Vehicles (BEV) and Plug-in Hybrid Electric Vehicles (PHEV) as two distinct portfolio options, and mapped by regions and OEMs by battery size and weighted average sales.

We use time-series disaggregated data from EV-volumes for this analysis over 4 years (ranging from 2019-2022), as well as refer to OEM-specific documentation including annual reports and product specification manuals. The three OEMs included in this analysis are Tesla [4], Toyota [5] and the Renault-Nissan-Mitsubishi (R-N-M) Alliance [6]–[8]. The overall approach for analysis is summarized in Fig. 1.

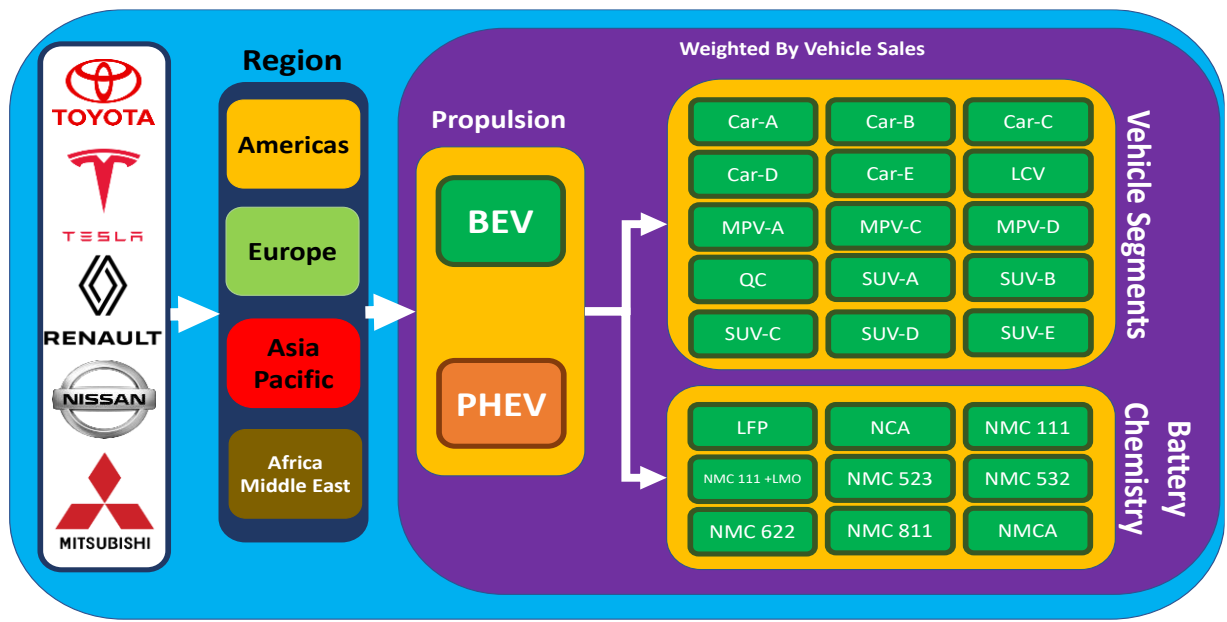

Figure 1: Proposed Framework

## 2 Regional Analysis

In this section, we provide insights on the trends in electrification by LDV segments as well as battery chemistry choices by BEV and PHEV, across regions.

### 2.1 Europe

In terms of LDV BEV segments for Europe, we find that in 2019, Car-D was the dominant segment in sales terms, whereas in 2020, Car-B emerges as the largest segment closely followed by Car-D. In 2021, Car-D becomes the largest segment with a 40% share, followed by Car-B (21%) and SUV-D (15%). In 2022, we see a shift to SUV-D which has a 54% share, followed by Car-D and Car-C (Fig. 3). The SUV-D segment sales increase about five fold between 2021 and 2022.

In terms of the battery size, the SUV-D segment has remained largely unchanged, going from 70 kWh in 2021 to about 72 kWh in 2022. In the Car-C segment, the average BEV battery pack size increases from 50 kWh in 2019 to about 57 kWh in 2022, whereas in Car-D, the average pack size increases from 75 kWh to 82 kWh. Also, electric LCVs have seen consistent sales over the years, with the battery pack sizes increasing from 36 kWh in 2019 to 48 kWh in 2022.

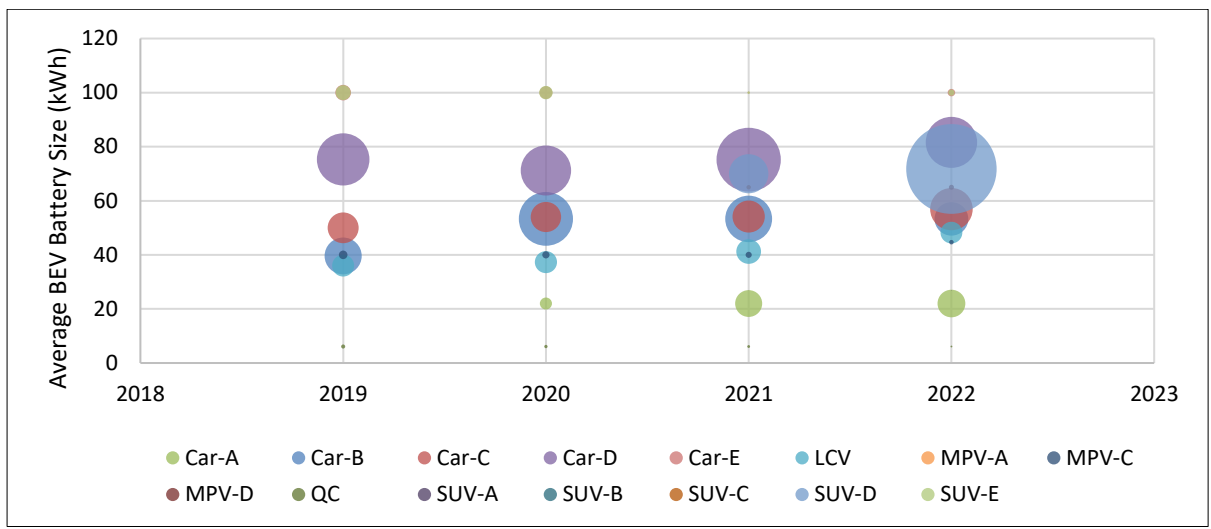

Figure 3: BEV LDV segments by battery size, weighted by sales for Europe (2019-2022)

In the case of LDV PHEVs, we find a shift from SUV-D in 2019 and 2020 to SUV-C in 2021 and 2022, as the dominant segments for the respective years. Interestingly, the PHEV SUV-C segment has been the fastest growing, with a 10-fold growth in sales volumes between 2020 and 2022. In the case of PHEVs, there is a clear trend towards SUVs, with SUV-C and SUV-B contributing to a majority of the sales (Fig. 4).

In terms of the battery size, the SUV-C segment has seen a decrease in average pack size, from 17.5 kWh to about 16.3 kWh, whereas in the case of SUV-B, it has remained more or less stable at 9.8 kWh.

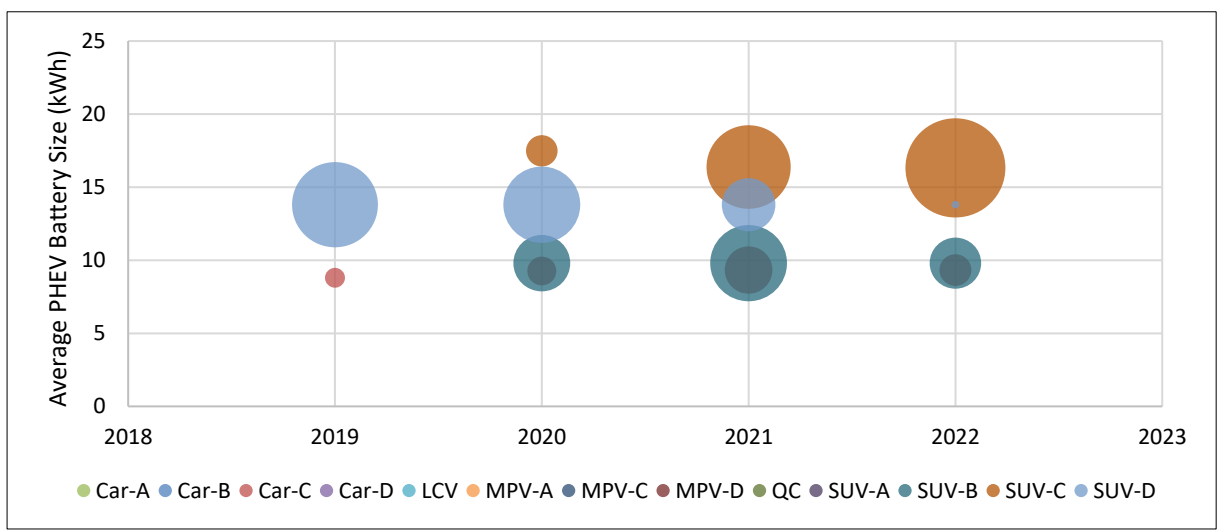

Figure 4: PHEV LDV segments by battery size, weighted by sales for Europe (2019-2022)

In terms of battery chemistries for BEVs (Fig. 5), NCA remains the dominant chemistry over the years, but with the entire demand coming from only Tesla. We also find that both NMC 523 and NMC 622 cells were used by OEMs in 2019, with the former delivering average battery packs of 30 kWh and the latter having an average pack size of 41 kWh. In 2020, with the exception of NCA demand from Tesla, the market pivoted almost entirely to NMC 622 cells, with average battery pack of 46 kWh. In 2021, again with the exception of NCA from Tesla, while NMC 622 remained the dominant cell chemistry of choice, we find a resurgence of NMC 523, but also the introduction of LFP (also used by Tesla, with the 60 kWh battery pack using LFP and the 86 kWh battery pack using NCA). In 2022, while NCA remains the largest share, we find LFP gaining significant momentum reaching 26% (compared to 7.5% in 2021). NMC 622 remains the other major chemistry at 22%.

In case of NMC 622, we find the average pack size increasing from 41 kWh in 2019 to 56 kWh in 2022. While NMC 523 declines in relative share by 2022, in absolute terms it remains relatively constant in 2021 and 2022, but being deployed for smaller battery packs ranging from 15-18 kWh.

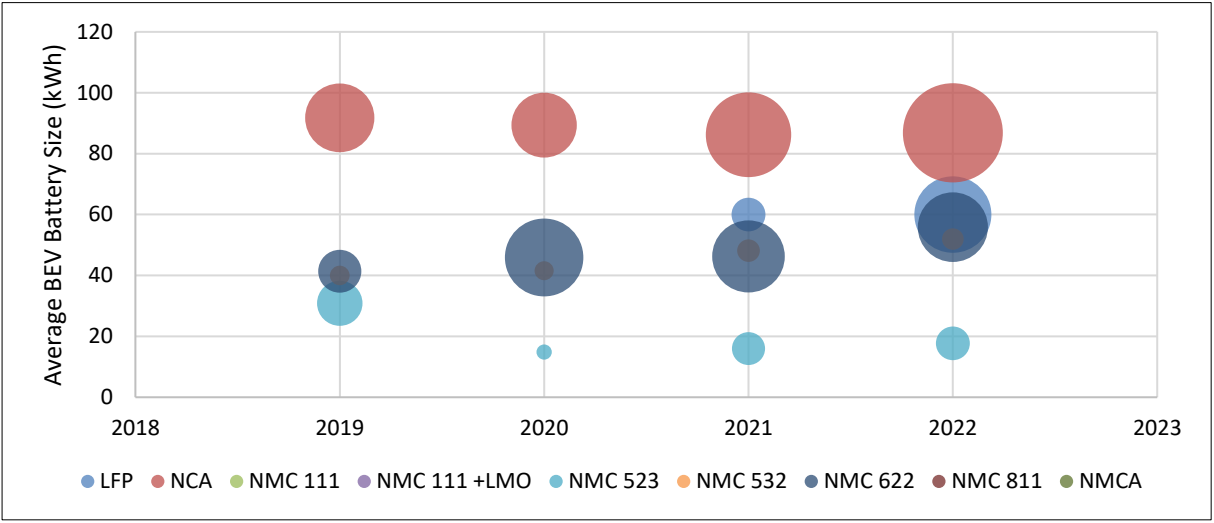

Figure 5: BEV LDV battery chemistries by battery size, weighted by sales for Europe (2019-2022)

In case of PHEVs (Fig. 6), we find that NMC 111 combined with LMO plays a key role across all four years, with an average battery pack size of 13.8 kWh. Between 2020-2022, NMC 523 also plays a key role as a battery chemistry choice (emerging as the largest share in 2021), with an average battery pack size of 9.8 kWh. In 2022, NMC 622 gains market share over NMC 523, with an average battery pack size of 18.1 kWh.

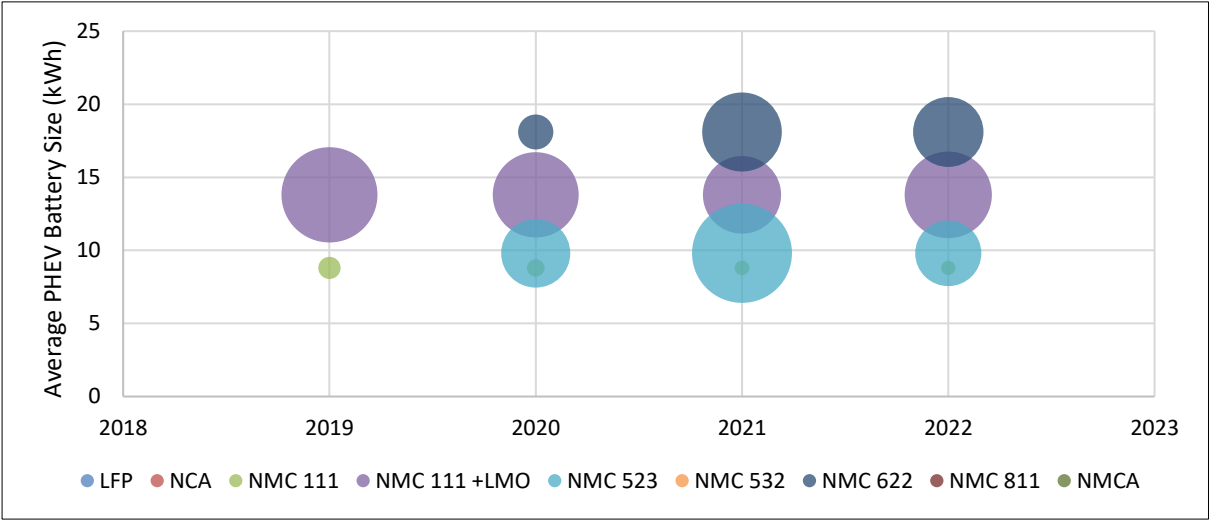

Figure 6: PHEV LDV battery chemistries by battery size, weighted by sales for Europe (2019-2022)

## 2.2 Americas

In terms of LDV BEV segments for the Americas region (Fig. 7), Car-D is the dominant segment in 2019 and 2020. In 2021 and 2022, the SUV-D segment becomes the largest share, achieving around 2/3$^{rd}$ in each year. This segment witnesses a 7-fold growth in 2022 compared to 2019. Although the relative share of Car-D decreases, in absolute terms the sales increase at around 11% CAGR between 2019-2022. Although from a low base, we also find that Car-E and SUV-E have a steady growth in volumes in the same period.

In terms of the battery size, the SUV-D segment goes from an average pack size of 80 kWh in 2020 to 70 kWh in 2022 (due to the additional offering of 60 kWh LFP variant by Tesla). In the Car-D segment, the average BEV battery pack size increases from 75 kWh in 2019 to about 81 kWh in 2022.

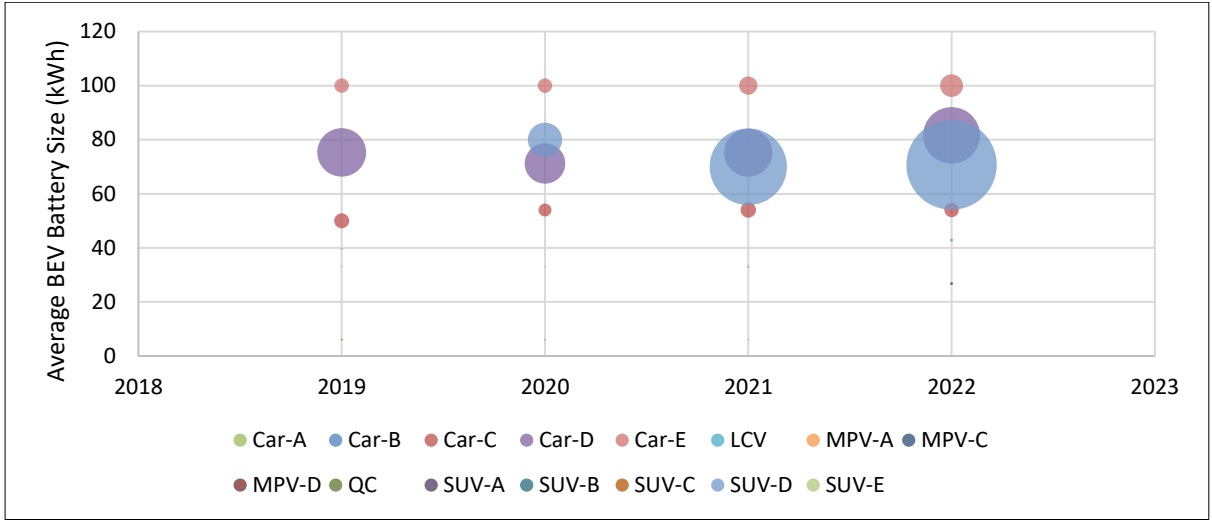

Figure 7: BEV LDV segments by battery size, weighted by sales for Americas (2019-2022)

In case of PHEVs, the SUV-D segment remains the dominant segment between 2019-2021. In 2022, the SUV-C segment becomes the leading segment (Fig. 8).

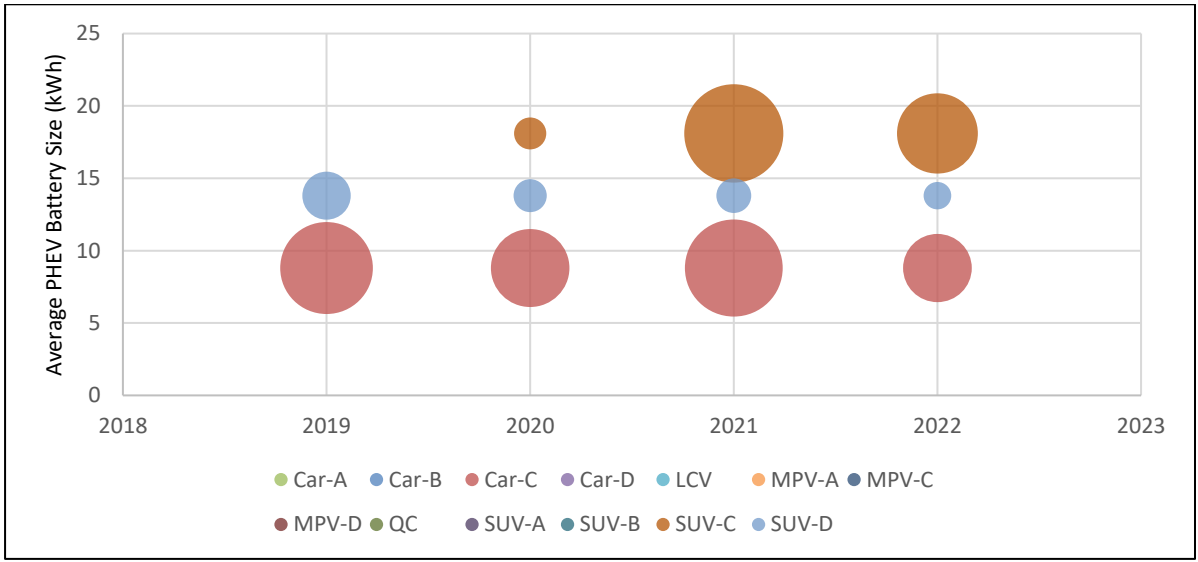

Figure 8: PHEV LDV segments by battery size, weighted by sales for Americas (2019-2022)

In case of BEV battery chemistries (Fig. 8), we find that NCA continues to remain the dominant chemistry choice, driven by Tesla's leading market share. In 2021 and 2022, LFP emerges as the second-most chemistry choice, also led by Tesla. In case of NCA, the average pack size comes down from 93 kWh to 87 kWh, due to Tesla offering two variants in the segment.

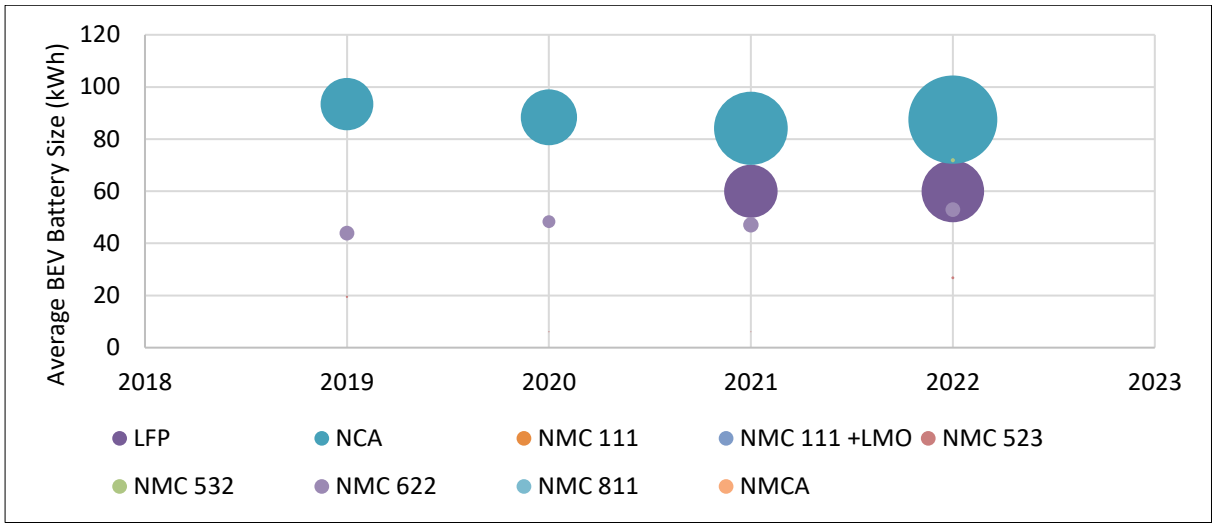

Figure 9: BEV LDV segments by battery chemistries, weighted by sales for Americas (2019-2022)

In case of PHEVs, the dominant chemistry choice from 2019 – 2021 is NMC 111 in combination with LMO, with the average pack size being 13.8 kWh (Fig. 10). In 2022, NMC 622 becomes the dominant chemistry choice, with the average pack size of 18.1 kWh.

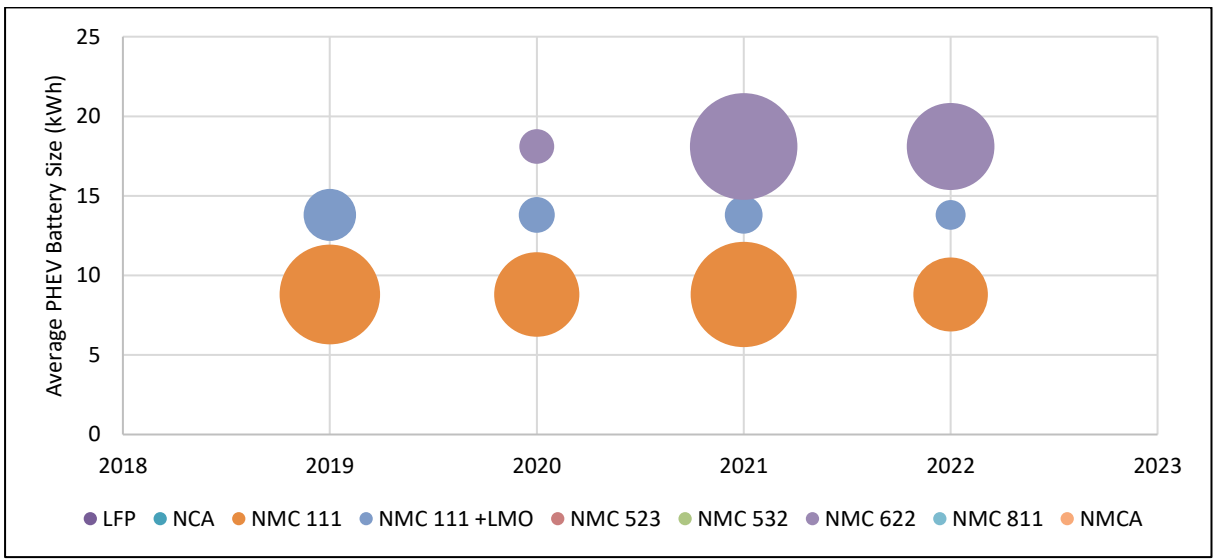

Figure 10: PHEV LDV segments by battery chemistries, weighted by sales for Americas (2019-2022)

## 2.3 Asia Pacific

In the Asia Pacific region, overall volumes are largely driven by China. In 2019, both Car-D and Car-C are the primary segments with regards to BEV sales (Fig. 11). In 2020, Car-D becomes the leading segment by far, with a market share of 86%. In 2021, both Car-D and SUV-D contribute about 48% each, with SUV-D doubling in share and sales volumes in 2022 to become the leading segment.

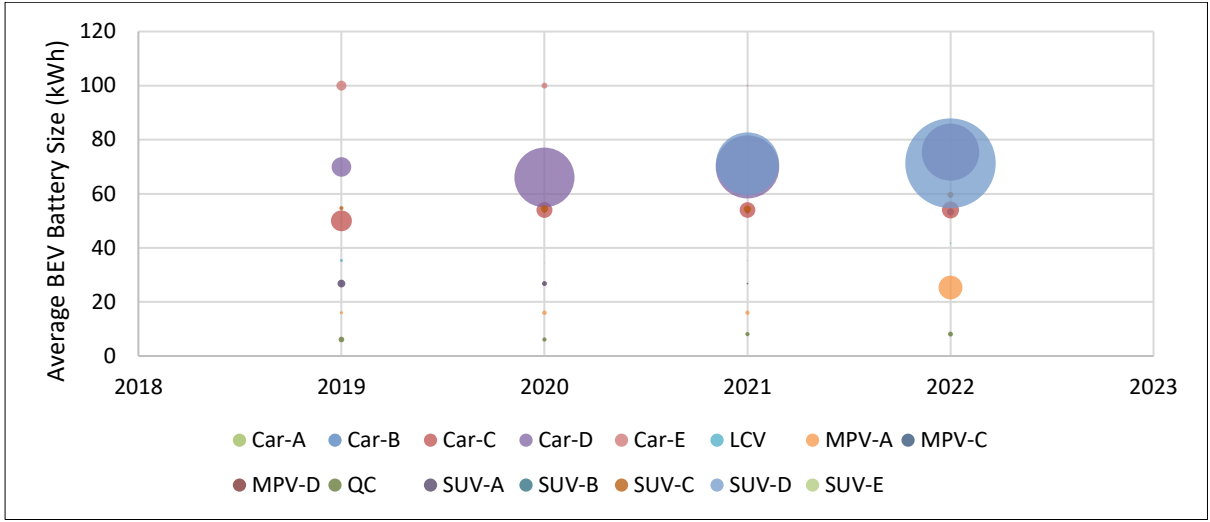

Figure11: BEV LDV segments by battery size, weighted by sales for Asia Pacific (2019-2022)

In case of PHEVs (Fig. 12), we see a transitional shift in segment choices, going from Car-C in 2019 to SUV-C in 2021, and then to both SUV-D and SUV-C in 2022. The average battery pack size is around 9.3 kWh for Car-C, around 16 kWh for SUV-C and around 14.5 kWh for SUV-D.

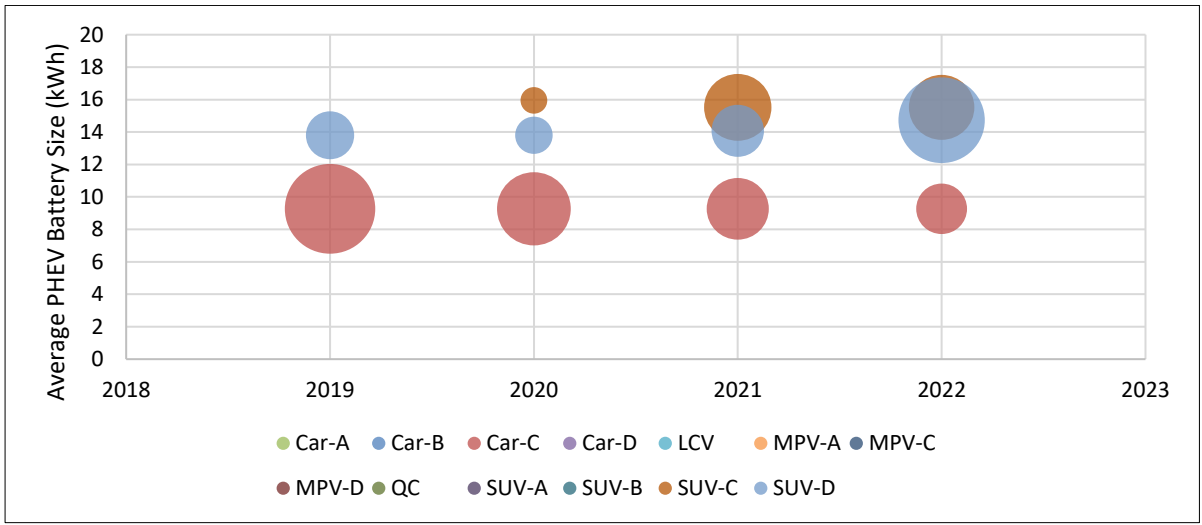

Figure12: PHEV LDV segments by battery size, weighted by sales for Asia Pacific (2019-2022)

In case of battery chemistries for BEVs ((Fig. 13), we find that in 2019, NMC 622 is the largest share followed by NCA. In 2020, LFP emerges as the largest share and continues to play a key role in subsequent years with a steady demand. In 2021 and 2022, we see the emergence of NMCA as the significant share in chemistries, essentially driven by Tesla's chemistry choice for its vehicles sold in China.

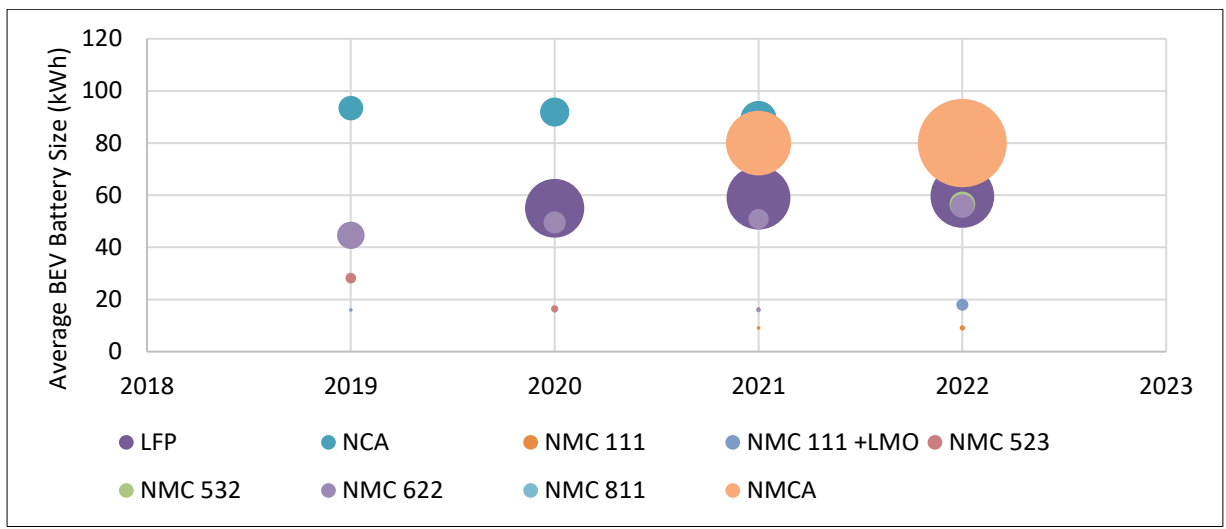

Figure 13: BEV LDV segments by battery chemistry, weighted by sales for Asia Pacific (2019-2022)

In the case of PHEVs (Fig. 14), NMC 111 remains the dominant base chemistry. While pure NMC 111 holds the largest share in 2019 and 2020, the NMC 111 combination with LMO continues to grow, resulting in the largest share in 2022.

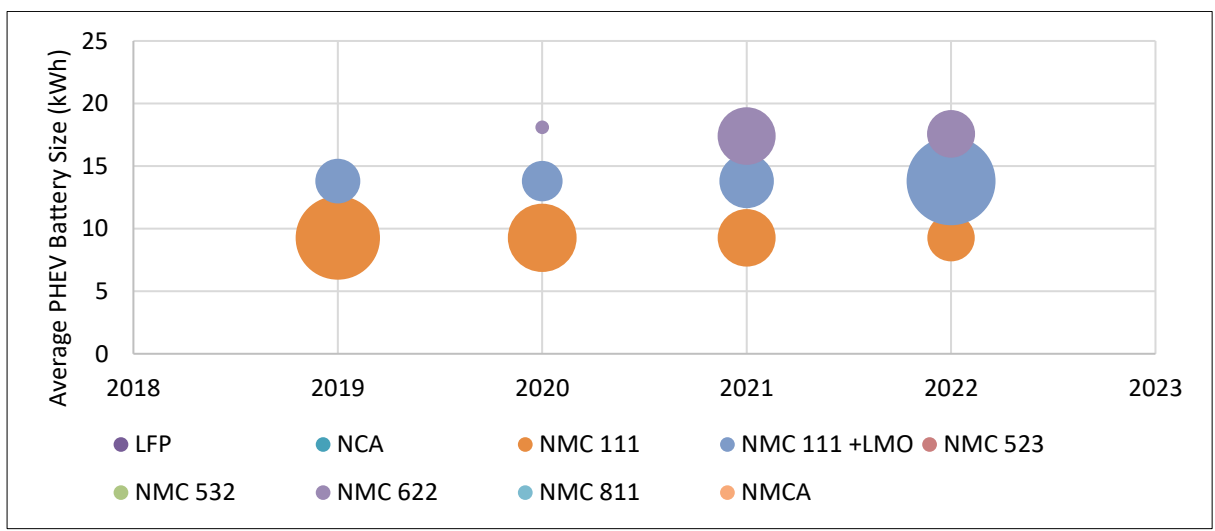

Figure 14: PHEV LDV segments by battery chemistry, weighted by sales for Asia Pacific (2019-2022)

## 2.4 Africa and Middle East

In the Africa and Middle East region, for BEVs, we find that in 2019 Car-B remains the dominant segment given low volumes (Fig. 15). In 2020 and 2021, Car-D emerges as the leading segment, whereas in 2022, although Car-D maintains a strong share, SUV-D emerges as the leading segment. For Car-B, the battery pack size increases from 40 kWh in 2019 to 53 kWh between 2020-2022. For Car-D, the average pack size increases from 71 kWh in 2020 to 81 kWh in 2022. In case of SUV-D, the pack size is around 70 kWh.

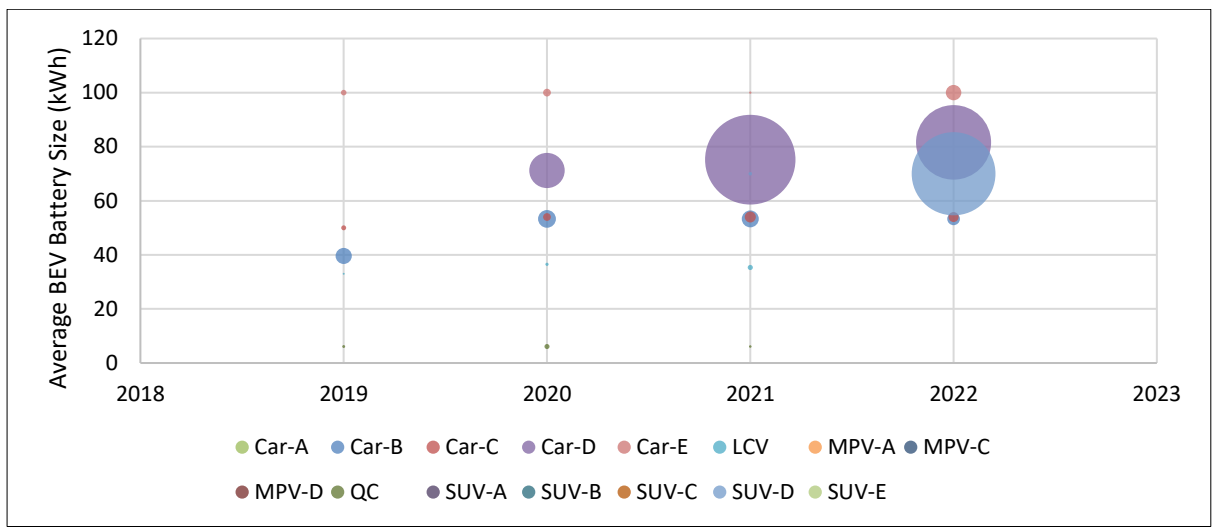

Figure15: BEV LDV segments by battery size, weighted by sales for Africa and Middle East (2019-2022)

In the case of PHEVs (Fig. 16), Car-C is the leading segment in 2019 and 2020. In 2021, we find that SUV-C and Car-C have equal shares, and in 2022, SUV-C surpasses Car-C to be the leading segment. In case of Car-C, the average pack size is 8.8 kWh, for SUV-C it is around 18.1 kWh and for SUV-D it is around 13.8 kWh.

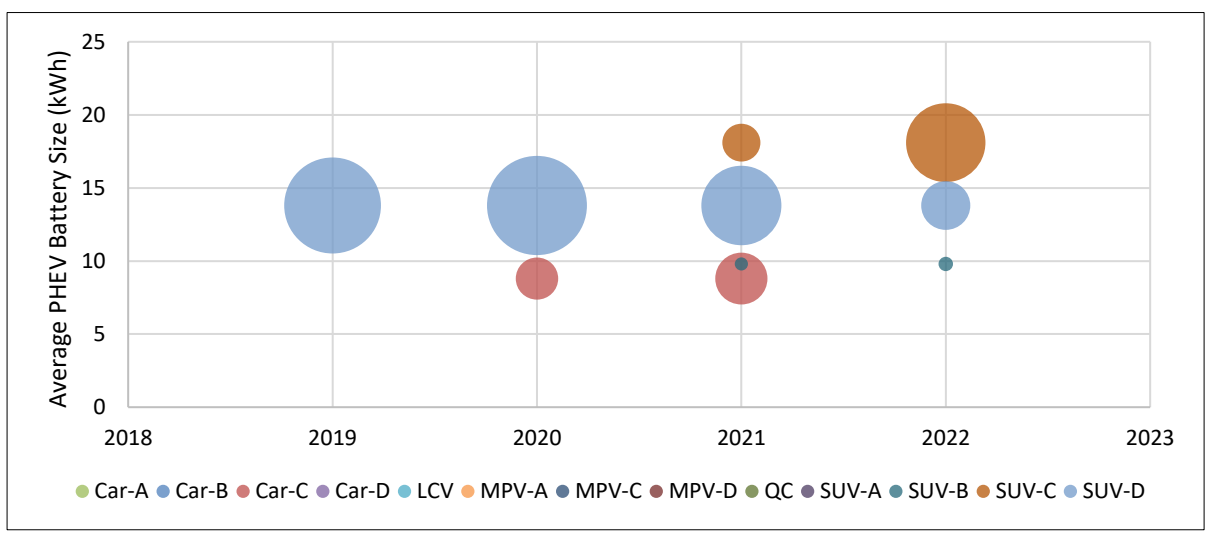

Figure16: PHEV LDV segments by battery size, weighted by sales for Africa and Middle East (2019-2022)

In case of battery chemistries for BEV (Fig. 17), NMC 523 is the main chemistry choice in 2019. In the following years, NCA emerges as the leading chemistry choice, with LFP also emerging in 2022. In 2019, NMC 523 has an average pack size of 26.2 kWh, after which it is used for smaller pack sizes of around 6 kWh upto 2021, and the phased out. Interestingly, in case of NCA, the average pack size remains constant for all models of Tesla, except for Model 3, wherein it increases from 72 kWh in 2019 to 81 kWh in 2022.

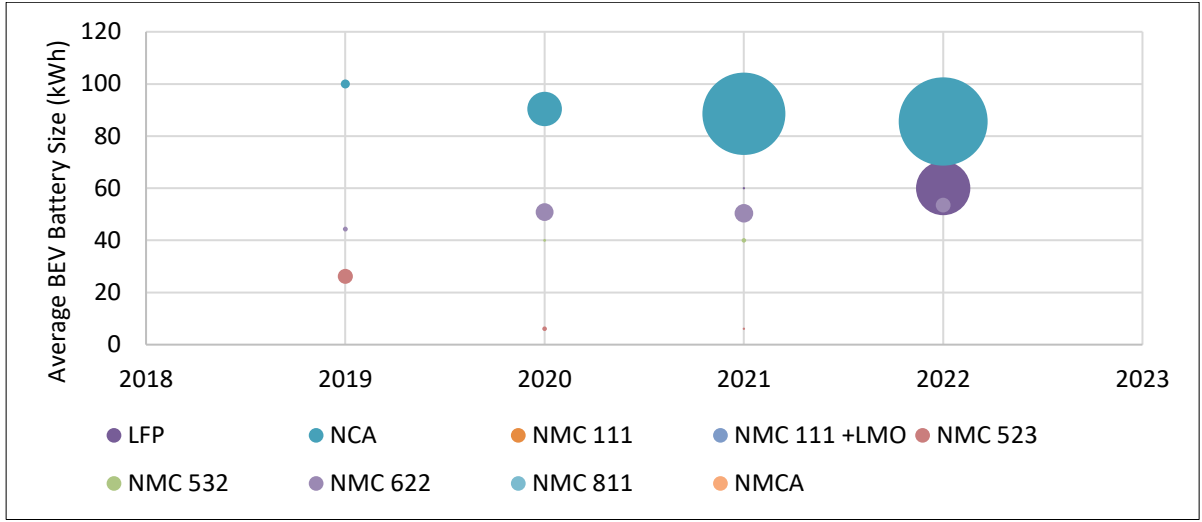

Figure17: BEV LDV segments by battery chemistry, weighted by sales for Africa and Middle East (2019-2022)

In case of PHEVs (Fig. 18), NMC 111 remained the primary chemistry in 2019 and 2020. NMC 622 emerges as a major choice in 2021, having an equal share with NMC 111. In 2022, NMC 622 becomes the largest share. The NMC 111 is primarily seen to be used in Car-C, while NMC 622 is used in SUV-C.

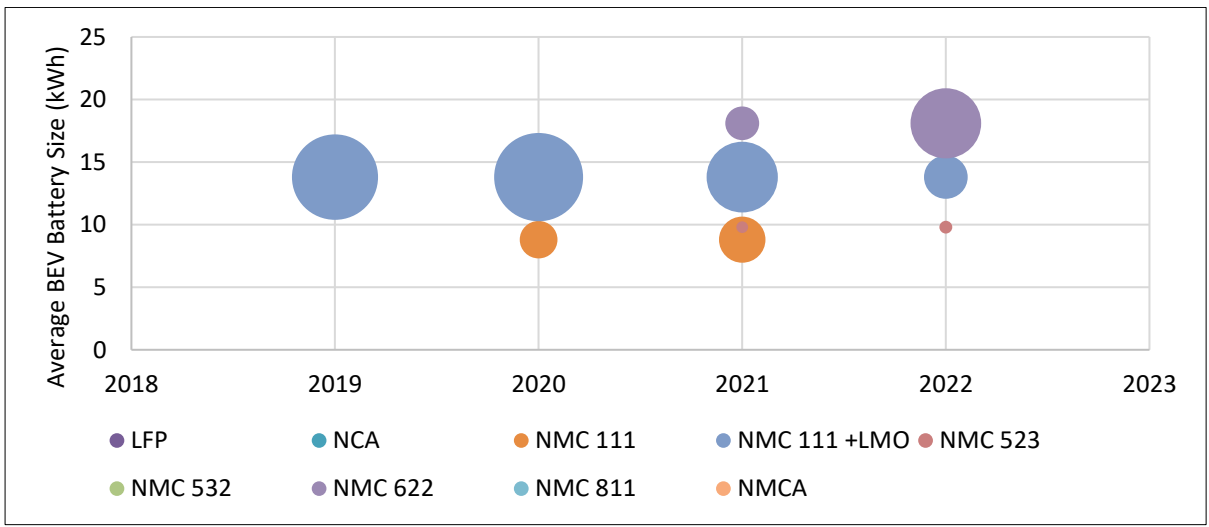

Figure18: PHEV LDV segments by battery chemistry, weighted by sales for Africa and Middle East (2019-2022)

## 3 Discussion and Conclusion

If we look at the summary of regional differences for the three OEMs analyzed in this paper, there is trend towards SUVs from Cars that is clear across both BEVs and PHEVs. Further, in the case of PHEVs, we find that SUV-C remains the dominant segment in 2022 across regions, compared to SUV-D in the case of BEVs.

We find that for BEVs, given the high market share of Tesla in all regions, the dominant chemistry choices are reflective of Tesla's strategy. We find that the dominant chemistry in 2022 is NCA for all regions, except in the Asia Pacific, where Tesla uses NMCA. In the Asia Pacific and Africa & Middle East, we find that in 2019, given the lower EV sales and entry-level models, the dominant chemistry was NMC 622 and NMC 523, respectively (Table 1). More importantly, average battery sizes for BEVs remain more or less similar in Europe and the Americas between 2019 and 2022, whereas in the Asia Pacific and Africa & Middle East, the 2019 battery sizes were much lower given the need for lower-cost EVs in these regions.

For PHEVs, the average battery sizes have increased across all regions, also driven by the regulatory need for a higher all-electric range for compliance with new emission standards. In terms of the battery chemistries, we find that NMC 111 in combination with LMO is prevalent across regions in 2019, with the exception of Asia Pacific, where NMC 111 is used, largely for cost reasons. In 2022, Europe continues with NMC 111 in combination with LMO, while Asia Pacific also migrates to the same. In the Americas and Africa & Middle East, we find a shift to NMC 622 in 2022 (Table 2).

Table 1: Regional comparison of trends in BEV segment, battery size and chemistry (2019-2022)

| Region | Dominant vehicle segment | | Sales wt. avg. battery size of dominant segment | | Dominant Chemistry | |
|---|---|---|---|---|---|---|
| | 2019 | 2022 | 2019 | 2022 | 2019 | 2022 |
| Europe | Car-D | SUV-D | 75 | 71 | NCA | NCA |
| Americas | Car-D | SUV-D | 75 | 71 | NCA | NCA |
| Asia Pacific | Car-C | SUV-D | 50 | 71 | NMC 622 | NMCA |
| Africa & Middle East | Car-B | SUV-D | 40 | 70 | NMC 523 | NCA |

Table 2: Regional comparison of trends in PHEV segment, battery size and chemistry (2019-2022)

| Region | Dominant vehicle segment | | Sales wt. avg. battery size of dominant segment | | Dominant Chemistry | |
|---|---|---|---|---|---|---|
| | 2019 | 2022 | 2019 | 2022 | 2019 | 2022 |
| Europe | SUV-D | SUV-C | 13.1 | 16.3 | NMC 111 + LMO | NMC 111 + LMO |
| Americas | SUV-D | SUV-C | 13.8 | 18.1 | NMC 111 + LMO | NMC 622 |
| Asia Pacific | Car-C | SUV-D | 9.2 | 14.7 | NMC 111 | NMC 111 + LMO |
| Africa & Middle East | SUV-D | SUV-C | 13.8 | 18.1 | NMC 111 + LMO | NMC 622 |

While this analysis presents a constrained perspective of three OEM choices, as the next steps, it is envisioned to expand this analysis to all global OEMs and apply the above methodology. This will help better understand regional differences in how the market responds to consumer choices with different technology pathways.